\documentclass{sig-alternate-05-2015}
\usepackage{listings}
\usepackage{color} 
\usepackage{float}

\definecolor{dkgreen}{rgb}{0,0.6,0}
\definecolor{gray}{rgb}{0.5,0.5,0.5}
\definecolor{mauve}{rgb}{0.58,0,0.82}

\begin{document}






%

\title{Optimizations and Heuristics to improve Compression in Columnar Database Systems}

%
%
%
%
%

\numberofauthors{2} 
%
\author{
%
%
\alignauthor
Jayanth\\
       \affaddr{Stanford University, CA, US}\\
       \email{jayanthj@stanford.edu}
}

\maketitle
\begin{abstract}
In-memory columnar databases have become mainstream over the last decade and have vastly improved the fast processing of large volumes of data through multi-core parallelism and in-memory compression thereby eliminating the usual bottlenecks associated with disk-based databases. For scenarios, where the data volume grows into terabytes and petabytes, keeping all the data in memory is exorbitantly expensive. Hence, the data is compressed efficiently using different algorithms to exploit the multi-core parallelization technologies for query processing. Several compression methods are studied for compressing the column array, post Dictionary Encoding. In this paper, we will present two novel optimizations in compression techniques - Block Size Optimized Cluster Encoding and Block Size Optimized Indirect Encoding - which perform better than their predecessors. In the end, we also propose heuristics to choose the best encoding amongst common compression schemes.
\end{abstract}

%
%


%
%

%
%


\keywords{Database Compression; Database Optimization; Encoding Schemes; In-memory Columnar Databases}

\section{Introduction}
Traditional database systems have relied on compression \cite{cormack1985data} mechanisms to improve performance significantly. Compression not just reduces the size of the data but also improves I/O performance. Columnar databases have gained momentum over the last decade \cite{abadi2006integrating}. In this paper, we discuss various state-of-art compression schemes in columnar database systems. We, further, propose optimizations and heuristics to arrive at the right compression scheme. A columnar database/column-store is one in which each attribute is stored in a separate column such that successive values are stored consecutively on the disk.  By virtue of storing successive values close, such databases present significant lossless compression ability in the form of dictionary encoding \cite{plattner2013course}\cite{plattner2012memory} . 

For any given column, distinct values appearing in the column are stored in a dictionary as an array with keys being the indexes of the array. The values comprising the dictionary are sorted. Each value in the sorted dictionary is identified by its \textbf{value ID} which is given by its position in the value array. The actual data in the column is stored as  an array of value IDs. The sorted dictionary and \textbf{value ID} array together represent the actual column.

Dictionary compression presents the advantage of executing search on the integer values. Since the dictionary is sorted by values, comparison and range selection queries can simply be performed on the value IDs. This yields better performance as opposed to when the actual values  stored in the columns are strings. For example, the query \textbf{\textit{select c from T where name > James}} can internally be replaced by finding all entries for which the value ID is greater than the value ID of the greatest value in the dictionary that is less or equal to \textit{James}. Furthermore in the context of in-memory columnar stores, dictionary compressed data can be loaded faster into the CPU cache. As the bottleneck is often data transport between memory and cache, performance gain exceeds the computing time needed for decompression.

\section{Compression Schemes}

Given dictionary encoding wherein a column gets encoded as a dictionary and value ID array, the size of the value ID array readily grows with large data volumes spanning billions of records. For in-memory systems, it is critical to further compress the value ID array using advanced compression schemes. These compression mechanisms are explained further \cite{plattner2012memory}.

\begin{figure*}
\centering
\includegraphics[width=\textwidth, height=0.45\textheight]{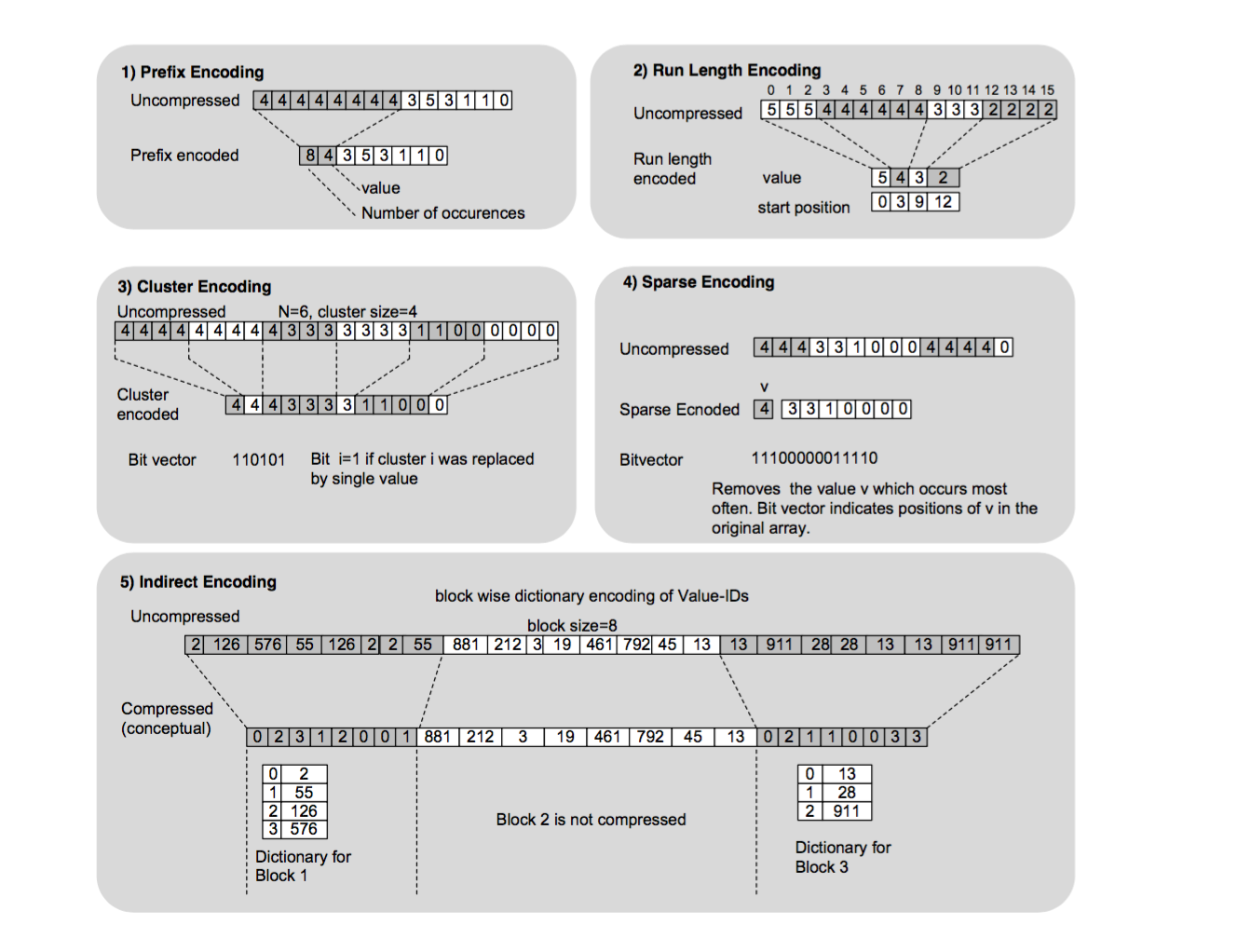}
\caption{Compression Schemes \cite{plattner2013course}}
\end{figure*}

\subsection{Prefix Encoding}
Real-world databases often comprise of columns that contain predominantly one value and remaining values are mostly unique. In order to apply prefix encoding, the value ID array is sorted such that this predominant value appears at the start of the array. The sequence is then replaced by storing the value once, along with the number of occurrences. 

\subsection{Run Length Encoding}
For a sorted value ID array, this encoding replaces the value sequences with a single instance of the value followed by the number of occurrences. 

\subsection{Sparse Encoding}
For value ID array containing predominantly a single value \textbf{V}, sparse encoding removes \textbf{V} and introduces a bit vector to specify all the positions at which V occurs in the original array. 

\subsection{Cluster Encoding}
Cluster encoding divides the value ID array into \textbf{N} fixed size blocks/clusters (typically 1024 elements). If a cluster contains unique value, it is replaced by the single instance of that value else it remains uncompressed. Additionally, a bit vector of length N indicates the clusters replaced by single value (1 if compressed else 0).

\subsection{Indirect Encoding}
Similar to cluster encoding, indirect encoding also addresses value ID array, divided into fixed size blocks. For each block containing few distinct values, an additional dictionary is used to encode the values in that block. There is a level of indirections introduced by local dictionaries for blocks in addition to a global dictionary.

\section{Optimisations} \label{sec:block}
While these advanced compression schemes prove useful, there are some basic flaws in their mechanisms. Both cluster and indirect encoding consider fixed size blocks. This does not provide optimal compression because it is highly unlikely that all elements in the cluster have the same value. Particularly, for in-memory systems it gets critical to fine tune the size of the blocks. The optimizations, discussed further, derive optimal block size based on the value ID array instead of using a fixed block size b ( 1024 elements, currently). This won't affect the query processor as instead of a fixed value 1024, it will use the block size specific to that column stored along with the value id array.

\subsection{Block Size Optimized Cluster Encoding}
Let \textbf{N} be the total size of the value ID array and \textbf{b} be the block size. ``size'' here represents number of elements constituting the entity. Without any compression, the size of the value ID array is given by \textit{N * size(valueID)}. Clustering encoding reduces the size to :
\begin{equation}
( N - (S*b)) * size(valueID) + S*size(valueID) + log(\frac{N}{b})
\end{equation}
where $S$ is the number of clusters, which contain only occurrences of a single value.
This equation can be re-written as:
\begin{equation}
(N - S*(b -1))  * size(valueID)
\end{equation}
considering log terms of lower complexity than rest, i.e. size of bit-vector is negligible. 
Thus, the optimal block size $b^{*}$ can be given mathematically as:
\begin{equation}
b^{*} = argmin _{b \in \cup_{i=1}^{log(N)} \{2^i\}}  (N - S*(b -1))
\end{equation}
Since, N is fixed for a dataset i.e. the count of records, it can removed to rewrite the optimization formultation as:
\begin{equation}
b^{*} = argmax _{b \in \cup_{i=1}^{log(N)} \{2^i\}}  S*(b -1)
\end{equation}

The above optimization algorithm can be solved iteratively in complexity of $O(Nlog(N))$. Further, optimization can be done to look for b till $\sqrt{N}$ and also, finding the value of S can be done dynamically from Run Length Encoding of the value ID array. Let, the value ID array is represented as $\cup_i  \{(v_i,c_i)\}$, where $v_i$ is the ith value ID and $c_i$ is its count. Let, $S_i$ denote the clustered blocks formed from the start of $(i-1)th$ value ID $v_{i-1}$ till the $ith$ value ID $v_i$ and $r_i$ denote the number of value IDs unclustered till start of $ith$ value ID.
Thus, $S_i$ and $r_i$ can be written as following recurrence relations, which can be exploited to solve using dynamic programming:

\begin{align}
S_1 &= 0 \\
r_1 &= 0 \\
S_i &= \frac{c_{i-1} - (|b - r_{i-1}| \mod b)}{b} \\
r_i &= (\sum_{j=1}^{i-1} c_j) \mod b  
\end{align}

Finally, the total number of clustered blocks $S$, required to solve the optimization problem can be derived as:
\begin{equation}
S = \sum_{i=1}^{..} S_i
\end{equation}

\begin{lstlisting}

def findClusteredBlocks(valCountArray,blockSize):
    S_i=[];R_i=[];S=0;R_i.append(0)
    for i in range(0,valCountArray.__len__()):
        R_i.append((R_i[i] + valCountArray[i][1]) % blockSize)
        S_i.append(max((valCountArray[i][1] - ((blockSize - R_i[i]) % blockSize))/blockSize, 0))
        S +=S_i[i]
    return S

def findOptimalBlockSize(valArray):
    valCountArray = convertValCount(valArray)
    Fmax = 0; OptimalBlockSize = 1
    for i in range(1, int(numpy.log2(valArray.__len__()))):
        b = 2 ** i
        S = findClusteredBlocks(valCountArray, b)
        F = S * (b - 1)
        print i, b, S, F
        if F > Fmax:
            Fmax = F; OptimalBlockSize = b
    return OptimalBlockSize
\end{lstlisting}

\subsection{Block Size Optimized Indirect Encoding}
In Indirect encoding, if a block contains only a few distinct values, an additional dictionary is used to encode the values in that block. The implementation also needs to store the information which blocks are encoded with an additional dictionary and the links to the additional dictionaries. If there are very many distinct values in a column, compression of the value ID array is not used. Indirect encoding can be applied efficiently, if data blocks hold a few distinct values as Indirect encoding saves space by mapping large global value IDs to small local value IDs.

The few distinct values criteria points towards usage of shanon entropy function, which has minima if all values are same while maxima, when all values are distinct and thus, have uniform distribution. However, each block should have as few distinct values and thus, on average the entropy of blocks should be minimized. Thus, the requirement for the blocks to have as few distinct values as possible can be used to formulate the following optimization problem for finding the optimal block size $b^{*}$:
\begin{equation}
b^{*} = argmin _{b \in \cup_{i=1}^{log(N)} \{2^i\}}  E[H_b(^bX_i)]
\end{equation} 
where $E$ is the expectation or average of the b-ary entropy $H_b(X_i)$ of the blocks $^bX_i$ defined by block size $b$. The reason for using b-ary entropy is to distinguish the entropy for differnt block sizes as the maximum different values the block can hold is $b$.
Here, the b-ary entropy $H_b(X_i)$ will be computed as:
\begin{equation}
H_b(X_i) = E[-log_b(P(X_i)]
\end{equation}

The above optimization problem can be solved using dynamic programming in complexity of $O(Nlog(N))$ by merging the frequency distribution at previous step into subsequent step.

\begin{lstlisting}
def findOptimalBlockSizeEntropy(valArray):
    Fmin = float("inf"); OptimalBlockSize = 1
    for i in range(1, int(numpy.log2(valArray.__len__()))):
        b = 2 ** i; data = []; S = 0
        for i in range(valArray.__len__()):
            data.append(valArray[i])
            if (i + 1) % b == 0:
                S += H(data, b)
                data = []
        F = S / math.ceil(valArray.__len__() / float(b))
        if F < Fmin:
            Fmin = F
            OptimalBlockSize = b
    return OptimalBlockSize
\end{lstlisting}

\section{Encoding decision heuristics} \label{sec:decision}
The choice between different encoding decisions is important, as wrong encoding sometimes may even lead to increase in data size, and thus, bad query performance.
Before proposing any algorithm or decision tree to decide over encoding decision, lets distinguish them on some important characteristics:

\begin{table}[H]
\centering
\caption{Heuristics Criteria} 
\begin{tabular}{|c||c|c|c|c|} \hline
Encoding & Blocks & Sorting & Sparsity & Prefix\\ 
\hline
Prefix & NO & * & YES & NO \\
Run-Length & NO & YES & NO & NO\\
Cluster & YES & * & NO & YES\\
Sparse & NO & NO & YES & YES\\
Indirect & YES & NO & NO & YES\\
\hline
\end{tabular}
\end{table}

Here, the columns in above table refer to the following:
\begin{enumerate}
\item \textbf{Blocks} - This column refers to the requirement of chunking the valueID array into blocks and thus, maximum block repetitions provide optimal compression. This holds as per the encoding algorithm only valid for Cluster and Indirect encoding. However, both encoding had an inherent limitation of using fixed block size as 1024 elements, which has been removed in section 3. 
\item \textbf{Sorting} - This column refers to the requirement of sorting the valueID array for good compression. Prefix encoding requires sorting the table in a way that puts the most frequent value at the beginning of the column. Cluster encoding, benefits from sorting but then, Run Length encoding is more optimal encoding choice.
\item \textbf{Sparsity} - This column refers to the requirement of sparsity in the valueID array, i.e. repetition of a single value. Apart from Sparse encoding, Prefix encoding also benefits from high sparsity, if the sparse value ID is the prefix itself. 
\item \textbf{Prefix Encoding} - This column refers to the possibility and benefit of applying prefix encoding before the encoding choice for more compression. Cluster encoding, indirect encoding and sparse encoding can be combined with prefix encoding, after sorting the frequent value ID as prefix.
\end{enumerate} 

\begin{figure}
\centering
\includegraphics[width=0.35\textwidth]{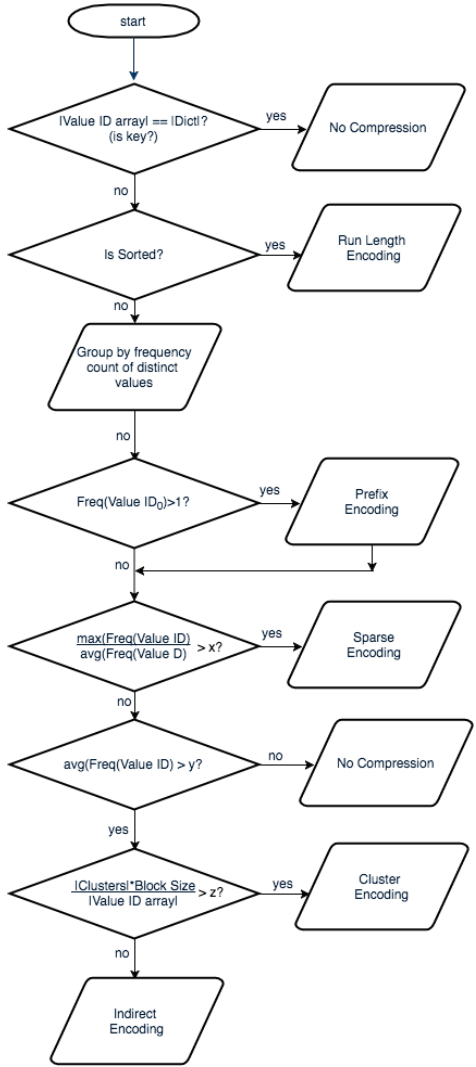}
\caption{Flowchart for encoding decision on a column}
\label{fig:decision}
\end{figure}

Based on these observations, a sample algorithmic flowchart for encoding decision can be proposed as figure \ref{fig:decision}.

The first condition checks whether the column is candidate key, i.e. all values are distinct. If all values are distinct, none of the above compressions will work. If all values are distinct, apply no compression. In special case, when the column is sequentially increasing or decreasing as in Time-Series table, linear Run Length Encoding should be applied. The next condition checks whether the column is sorted, i.e. all the value IDs are in order. 
If the value ID array is sorted, then Run Length Encoding should be applied as it gives the optimal compression for sorted value id array. Then, the heuristics finds the frequency of each distinct value ID in the value ID array using a group by distinct count query. If the count of 1st value ID in value ID array is greater than 2 and that values repeats itself in the beginning, then Prefix Encoding should be applied. 
The following condition checks whether the sparsity in the value ID array is greater than x, a tune-able parameter (x>1). The sparsity is quantified as the ratio of the maximum of the frequency of value ID array and the average of the frequencies of value ID array i.e. how much more is the most sparse value ID repeating over average repetition. If the sparsity in the value ID array is significantly large, then Sparse Encoding should be applied.
The next condition checks whether there is sufficient repetition in the value ID array i.e. the average repetition is greater than y, a tune-able parameter (y>1). The average repetition is quantified as the average of the frequencies of value ID array. 
The final condition checks whether the clustered value IDs after cluster encoding lead to high compression. The ratio of the size of cluster of compressed value IDs using cluster encoding and the total value ID array size is compared with z, a tune-able parameter (0<z<1).  Here, S is the number of the clustered blocks and b is the optimal block size, both values calculated as in in section 3 after applying the algorithm for calculating optimal block size in cluster encoding. If the clustered value IDs after cluster encoding lead to high compression, then Cluster Encoding should be applied. If the condition to apply cluster encoding is not met but there is sufficient repetition, then Indirect Encoding should be applied. 
\section{Conclusions}
Optimal block size in cluster and indirect encoding perform better than the conventional techniques which uses fixed block size. Furthermore, the heuristics based on certain tunable parameters can hint towards the right compression scheme. While dictionary encoding ensures good compression, optimizations and heuristics to choose the right advanced compression technique can further reduce the storage footprint.


%
\bibliographystyle{abbrv}
\bibliography{sigproc}  

\begin{thebibliography}{1}

\bibitem{abadi2006integrating}
D.~Abadi, S.~Madden, and M.~Ferreira.
\newblock Integrating compression and execution in column-oriented database
  systems.
\newblock In {\em Proceedings of the 2006 ACM SIGMOD international conference
  on Management of data}, pages 671--682. ACM, 2006.

\bibitem{cormack1985data}
G.~V. Cormack.
\newblock Data compression on a database system.
\newblock {\em Communications of the ACM}, 28(12):1336--1342, 1985.

\bibitem{plattner2013course}
H.~Plattner.
\newblock {\em A course in in-memory data management}.
\newblock Springer, 2013.

\bibitem{plattner2012memory}
H.~Plattner and A.~Zeier.
\newblock {\em In-memory data management: technology and applications}.
\newblock Springer Science \& Business Media, 2012.

\end{thebibliography}
%
%

\end{document}